\documentclass[aps,preprint,showkeys,
amsmath,amssymb,
aps,
]{revtex4-1}

\usepackage{graphicx}
\usepackage{dcolumn}
\usepackage{bm}
\usepackage{color}

\begin{document}

\date{\today}


\title{Partially controlling transient chaos in the Lorenz equations}

\author{Rub\'en Cape\'ans}
\email{ruben.capeans@urjc.es}
\affiliation{Departamento de F\'isica, Universidad Rey Juan Carlos, Tulip\'an s/n, 28933 M\'ostoles, Madrid, Spain}

\author{Juan Sabuco}
\email{juan.sabuco@urjc.es}
\affiliation{Departamento de F\'isica, Universidad Rey Juan Carlos, Tulip\'an s/n, 28933 M\'ostoles, Madrid, Spain}

\author{Miguel A. F. Sanju\'an}
\email{miguel.sanjuan@urjc.es}
\affiliation{Departamento de F\'isica, Universidad Rey Juan Carlos, Tulip\'an s/n, 28933 M\'ostoles, Madrid, Spain}

\author{James A. Yorke}
\email{yorke@umd.edu}
\affiliation{University of Maryland, College Park, Maryland 20742, USA}

\begin{abstract}
Transient chaos is a characteristic behavior in nonlinear dynamics where trajectories in a certain region of phase space behave chaotically for a while, before escaping to an external attractor. In some situations the escapes are highly undesirable, so that it would be necessary to avoid such a situation. In this paper we apply a control method known as \textit{partial control} that allows one to prevent the escapes of the trajectories to the external attractors, keeping the trajectories in the chaotic region forever. To illustrate how the method works, we have chosen the Lorenz system for a choice of parameters where transient chaos appears, as a paradigmatic example in nonlinear dynamics. We analyze three quite different ways to implement the method. First, we apply this method by building a 1D map using the successive maxima of one of the variables. Next, we implement it by building a 2D map through a Poincar\'{e} section. Finally, we built a 3D map, which has the advantage of using a fixed time interval between application of the control, which can be useful for practical applications.
\end{abstract}

\maketitle

\section{Introduction}

Traditionally, the aim of classical control methods in chaotic systems has been to lock the dynamics into a specific steady state or periodic orbit (see for instance \cite{ControlCha,ControlD,StabilityDelay}. But there has arisen a need for other approaches because there have appeared many situations where chaos could be a powerful attribute. In mechanics, for example, chaos helps prevent undesirable resonances  \cite{Oscillator}. In engineering, the thermal pulse combustor is more efficient in the chaotic regime \cite{Thermal}. In living organisms, chaotic dynamics in vital functions can make the difference between health and disease \cite{Perc}. In biology, it has been suggested that the disappearance of chaos may be the signal of pathological behavior \cite{Biological}. In all these cases, chaos is a desirable property that is worth preserving.

However, sometimes the chaotic behavior is only transient in nature, and it is necessary to apply external perturbations to keep trajectories in the transient chaotic regime. Transient chaos is a characteristic dynamical behavior that occurs in a certain region of phase space, where chaotic orbits exist for a while, before escaping to an external attractor. This kind of behavior can be found in a broad variety of systems like the periodically driven $CO_2$ laser \cite{Laser}, voltage collapse in electrical power systems \cite{Dhamala}, or the Mcann-Yodzis ecological model \cite{McCann}, among many others.

From a topological point of view, transient chaos is caused by the presence of a chaotic saddle in phase space. A chaotic saddle can arise as a parameter is varied, when a chaotic attractor collides with the boundary of its own basin of attraction, causing a \textbf{boundary crisis}. Then the chaotic attractor disappears, allowing the trajectories to escape to an external attractor. In many situations, the external attractor may be an undesirable dynamical state. For example, in the context of ecology \cite{Ecology}, the escape may result in the extinction of some species, while in the cancer model described in Ref.~\cite{Cancer}, the dynamics evolves towards a state where an undesirable growth of tumor cells occurs.

With the aim of avoiding the undesirable escapes, different control methods have been proposed in the literature \cite{Schwartz,Dhamala,Bertsekas,Bertsekasdos}. These methods have been mainly designed to be applied in deterministic systems. However, when we are implementing some control method in a real system, the presence of disturbances may be unavoidable and must be considered, especially when it is necessary to keep the control as small as possible. Methods that perform well in systems in absence of disturbances can fail dramatically when disturbances appear. For this reason, it is reasonable to consider a term, that we call \textbf{disturbance}, that encloses all the uncertainty affecting the dynamics of the system, like modeling mismatches, finite precision in the measure of initial conditions or even systematic or random external disturbances.

\begin{figure}
\includegraphics [trim=0cm 0cm 0cm 0cm, clip=true,width=0.55\textwidth]{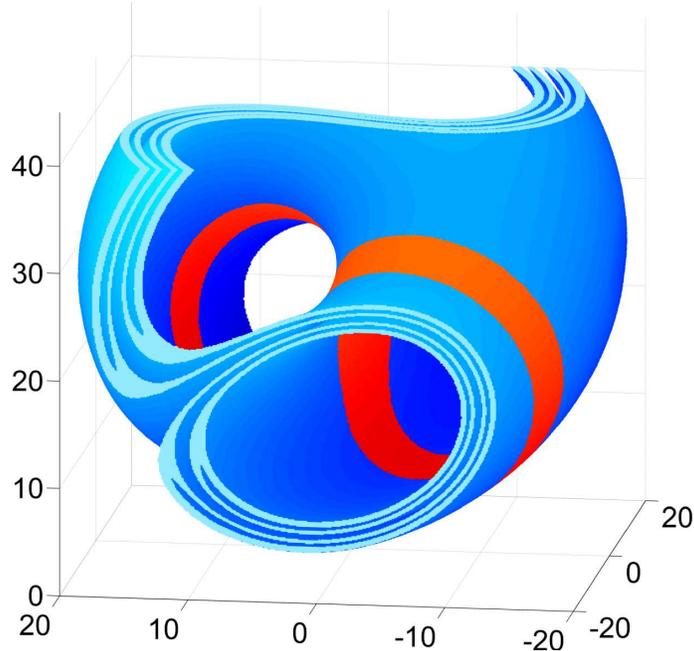}\\
\centering
\caption{\textbf{Example of the set needed to partially control the Lorenz system.} The figure shows an example of a set in the phase space computed for the partially controlled Lorenz system in the transient chaotic regime. The blue set represents the points of the phase space that satisfy the control condition defined by the partial control method. The red set is a subset of the blue set, and represent the asymptotic region where the controlled dynamics converges.}
\label{0}
\end{figure}

To reduce the amount of control necessary to avoid escapes in a transient chaotic system in the presence of disturbances, a control method called \textbf{partial control} has been proposed in Refs.~\cite{Automatic,Asymptotic}. This method is based on finding certain set in the phase space, which allows to avoid the escape of the trajectories. Indeed the control idea, based on controlled set invariance, is pretty standard (see Refs.~\cite{Invariant,Dhamala,Hutson,Genesio,Kolmanovski}), however no partial control works have appeared in the control literature. This method consider that the trajectories are affected by a bounded disturbance, and then a bounded control is applied. Both ideas, the control constraint and the application after a perturbation is not new in the literature (see for example \cite{Gutman,Gutmandos,Gutmantres,Astrom,Donkers}). Nevertheless the use of sets to control transient chaotic systems affected by disturbances remains unexplored until now. The shape of the invariant sets play an important role in the dynamics of the controlled system \cite{Nagumo,Invariant}. This situation is even stronger in the case of partial control, where the invariant set can be rather complex and it is only possible to find it using a numerical algorithm \cite{Automatic}. In Fig.~\ref{0} an example of a set computed for the Lorenz system is shown. The remarkable features of this method is that is able to use a control smaller than the disturbances affecting the system. In addition, the partially controlled dynamics remains chaotic, preserving the natural dynamics of the system. The method was successfully applied to several paradigmatic systems like the H\'{e}non map and the Duffing oscillator \cite{Automatic}, as well as other models in the context of ecology, cancer dynamics or economy \cite{Ecology,Cancer,Das}.

The partial control method is applied to maps, so that when we want to apply it to ordinary differential equations, a suitable time discretization of the continuous dynamics is needed to obtain a discrete time map. There exists a wide literature in Nonlinear Dynamics where the control is applied in a discrete way to continuous systems to suppress a chaotic behavior. For example, Refs.~\cite{Brain,ExpChaos} describe two different experimental setups where a continuous system is controlled using a discrete time controller. Different discretization techniques are possible, as for example, cutting the flow with a Poincar\'{e} section, or taking successive maxima (or minima) of the time series of a certain variable. Alternatively, it could also be useful in applying the control at certain predefined values of time, as in the case of medical treatments based on periodic interventions. In this sense, we propose a way to build this kind of map using a time-discretization technique. With this approach it will be possible to control the system with a fixed time interval, which can be an advantage in many real applications.

With the aim of showing how this method works in a flow affected by some disturbance, we have chosen the paradigmatic Lorenz system for a choice of parameters where transient chaos appears and escapes occur. To apply the control method, we consider three different ways to discretize the dynamics of the flow taking into account how the disturbance in the flow appears in the map. First, a 1D (one-dimensional) map is built taking successive maxima of one of the variables. Next, a 2D (two-dimensional) map is obtained from a Poincar\'{e} section. Finally, in the third case a 3D (three-dimensional) map is built from a time-discretization of the flow. In all these cases we show how the partial control method is implemented (the codes are available in Ref.~\cite{codes}). The procedure considered here can be applied in a similar way to a wide variety of systems found in the literature, where the goal is to avoid the escapes associated with a transient chaotic dynamics.

The structure of the paper is as follows. Section $2$ is devoted to a general description of the partial control method. In Section $3$, we apply the method to the Lorenz system, demonstrating the application of partial control in dimensions 1, 2 and 3, (paying special attention to the novel 3D case) and highlighting the pros and cons of the extra dimensions. Finally, some conclusions are drawn in Section $4$.

\section{A general description of the partial control method}

The partial control method is a recently developed control strategy for preventing escapes associated with a transient chaotic region in systems affected by disturbances. It is particulary appropriate when it is desirable to keep the magnitude of the control small.

 The method is based on the existence of certain sets, known as \emph{safe sets}, which are used for steering the trajectory with small controls so that escapes can be avoided. In addition, the chaotic behavior of the dynamics is preserved. This control method is applied on maps, so in the case of flows affected by disturbances, it is necessary to previously discretize the dynamics. We consider here, that the discrete dynamics can be modeled as $q_{n+1}=f(q_n)+\xi_n$ where $\xi_n$ is an additive term representing the disturbance, which we assume to be bounded by some $\xi_0$. For the controller, the observable is $[f(q_n)+\xi_n]$. The controller cannot measure $f(q_n)$ or $\xi_n$ separately in the real time application of the control.

In the partial control method the control variables are the same variables of the system. But we do not have a full control of what happens with those variables. We assume that we only apply a discrete control every $\Delta t$. To relate the control in the map with the control in the physical continuous time model, we assume that the control is applied almost instantly in the flow, that is, we assume that the time spent to perturb the trajectory is much lower than the typical time variation of the dynamics.

 The control scheme is $q_{n+1}=f(q_n)+\xi_n+u_n$, where $u_n$ represents the applied control that we also consider to be bounded by some $u_0$. One of the main achievements of this method is the relationship between the value of disturbance $\xi_0$ and the value of control $u_0$. If we have $u_0>\xi_0$, it would be trivial to have the control overpower the disturbance. However our goal is not to determine the trajectory, but only to prevent the escapes, and surprisingly, it is possible by using  $u_0<\xi_0$, which is rather counterintuitive.

\begin{figure}
\fboxsep=0mm
\includegraphics [trim=0cm 0cm 0cm 0cm, clip=true,width=0.62\textwidth]{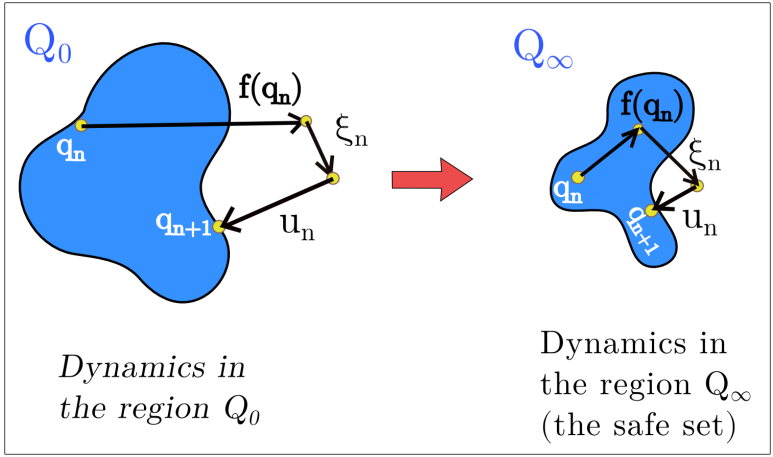}\\
\centering
\caption{\textbf{Dynamics in $Q_0$ and $Q_\infty$.} The left side shows an example of a region $Q_0$(in blue) in which we want to keep the dynamics described by $q_{n+1}=f(q_n)+\xi_n+u_n$.  We say that $|\xi_n| \leq \xi_0$ is a bounded disturbance affecting the map, and $u_n$ is the control chosen so that $q_{n+1}$ is again in $Q_0$. To apply the control, the controller only needs to measure the state of the disturbed system, that is $[f(q_n)+\xi_n]$. The knowledge of $f(q_n)$ or $\xi_n$ individually is not required. The right side of the figure, shows the region $Q_\infty \subset Q_0$ (in blue), called a \emph{safe set}, where each $x_n \in Q_\infty$ has $x_{n+1} \in Q_\infty$ for some control $|u_n|\leq u_0$, which is chosen depending on $\xi_n$. Notice that the removed region does not satisfy $|u_n|\leq u_0$.}
\label{1}
\end{figure}

To apply the method, we initially have to identify a region $Q$ in phase space with transient chaos. Trajectories in $Q$ follow the  chaotic dynamics and eventually escape from $Q$ to an external attractor. The goal is to keep the dynamics $q_{n+1}=f(q_n)+\xi_n$ within the region $Q$, and the partially controlled trajectories must satisfy the condition  $\xi_0>u_0 \geq |u| >0$. On the left side of Fig.~\ref{1}, we display an example of the dynamics in the region $Q_0 = Q$. Some points may need a big control to return to $Q_0$, and therefore we remove them to preserve only the set of points that need only a small control bigger than some selected $u_0$. Following this idea, it is possible to numerically find a limiting set $Q_\infty \subset Q$, where all the $q_n$ can be kept. In a formal way, we will say that $Q_\infty$ is a \textbf{safe set} for the specified $\xi_0$ and $u_0$, if for every $q \in Q_\infty$ and any $\xi$ where $|\xi| \leq \xi_0$, there is a $u$ with $|u| \leq u_0$ such that $f(q) +\xi+u \in Q_\infty$. The control $u_n$ is chosen with the knowledge of $f(q_n)+\xi_n$, and applied to place the trajectory again in the set $Q_\infty$. We say that trajectories found under these conditions are \textbf{admissible trajectories}. Sometimes, the set $Q_\infty$ can consist of many components, while others is a connected set like the right side of Fig.~\ref{1}, where we also have shown a partially controlled trajectory.

\begin{figure}
\includegraphics [trim=3cm 0cm 0cm 0cm, clip=true,width=0.73\textwidth]{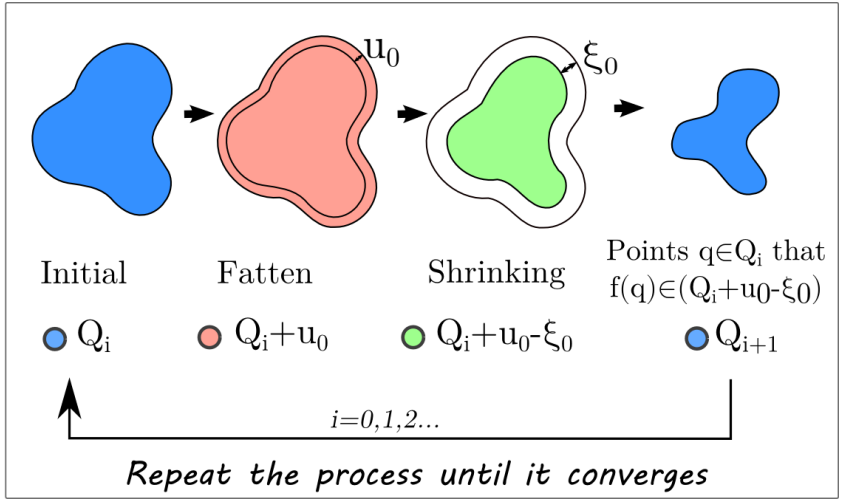}\\
\centering
\caption{\textbf{Graphical process used by the Sculpting Algorithm to obtain the safe set.} The denoted set $Q_i$ is fattened by the thickness $u_0$. The fattened set is displayed in red. Then, the new set is shrunk or contracted by a distance $\xi_0$, obtaining the set denoted $Q_i+u_0-\xi_0$ (in green). Finally we remove the grid points $q \in Q_i$ whose image $f(q)$ falls outside $Q_i+u_0-\xi_0$. Notice that $Q_{i+1} \subset Q_i$.}
\label{2}
\end{figure}

One of the advantages of this method is that the set $Q_\infty$ can be determined computationally following an iterative process. The set $Q$ is represented by a grid stored in a computer. Beginning with the region $Q_0=Q$, in the first iteration we remove the grid points $q \in Q_0$ for which there are $\xi$ with $|\xi| \leq \xi_0$ such that $f(q)+\xi$ cannot be moved back inside $Q_0$ using a $u$ for which $|u| \leq u_0$. As a result of this first pruning, a new region $Q_1\subset Q_0$ is obtained. Applying the same process to $Q_1$, we obtain a smaller set $Q_2 \subset Q_1 \subset Q_0$. Repeating this process until it converges, the final set denoted $Q_\infty$ is found. This set is known as the \textbf{safe set}. Based on this idea, we create an algorithm  that we called the \textbf{Sculpting Algorithm} \cite{Automatic}, for computing the successive regions $Q_n$ until the safe set is finally found. We illustrate the procedure of finding the safe set in Fig.~\ref{2}. We are given the bound $u_0$ and $\xi_0$ and the region $Q_0=Q$. The $i^{th}$ step can be summarized as follows:

\begin{enumerate}
  \item  Fatten the set $Q_i$ by $u_0$, obtaining the set denoted $Q_i+u_0$.
  \item  Shrink the set $Q_i+u_0$ by $\xi_0$, obtaining the set denoted $Q_i+u_0-\xi_0$.
  \item  Let $Q_{i+1}$ be the points $q$ of $Q_i$, for which $f(q)$ is inside the set denoted $Q_i+u_0-\xi_0$.
  \item  Return to step $1$, unless $Q_{i+1}=Q_i$, in which case we set $Q_\infty=Q_i$. We call this final region, the \textit{safe set}. Note that if the chosen $u_0$ is too small, then $Q_\infty$ may be the empty set and a bigger value of $u_0$ must be selected.
\end{enumerate}

To implement the algorithm, we need to choose a grid of points in which we represent the set $Q_i$. As we remove points from the grid representation of $Q_i$, the process eventually stops when $Q_{i+1}=Q_i$ for some $i$, and we write $Q_\infty=Q_i$ for that $i$. Due to the complex shape of the chaotic saddle underlying the chaotic dynamics, the derivation of a rigorous proof of the convergence of the algorithm would be extremely difficult. However we can show in a very intuitive way that the algorithm converges in a finite number of steps to a safe set. To find the safe set, we begin with a grid of points covering $Q$ which contains a finite number of points. Then, the Sculpting Algorithm removes in each iteration the points that do not satisfy the control condition. As a result, subsets $Q_n\subset... Q_2 \subset Q_1 \subset Q$ are obtained. We iterate this process until $Q_{n+1}=Q_n$, being $Q_n$ the safe set. Therefore the finite number of initial points of $Q$ ensures that the iterative process converges to a safe set, if it exists, in a finite number of steps.


Finally it is important to mention the influence of the grid resolution. The finite resolution of the grid implies a certain imprecision in obtaining the safe set. If we call $r_j$ the grid resolution spacing in each dimension, the total error (the maximum distance to the nearest grid point) will be  $\sqrt{\sum (r_j/2)^2}$. For example, if we have a 2D grid with resolution $r_x=r_y=0.001$, the error in the representation of a point $q_n$ will be $0.001/\sqrt{2}$. For a good precision in the computation, we recommend here to take a grid resolution $10$ times smaller than the magnitude of the control $u_0$.  With this resolution the shape of the safe set usually remains practically unchanged with respect to the safe sets obtained with higher resolution. This practical recommendation gives a relative error in the control of $5\%$, that is, when we are applying control we will have to put only a $5\%$ more of control in the worst cases. The increase of the resolution grid improves the precision, but the computational time has a polynomial growth with the dimension of the map, so there is a trade-off between the precision and the computational cost.

\section{Avoiding escapes in the Lorenz system}

\begin{figure}
\includegraphics [clip=true,width=0.9\textwidth]{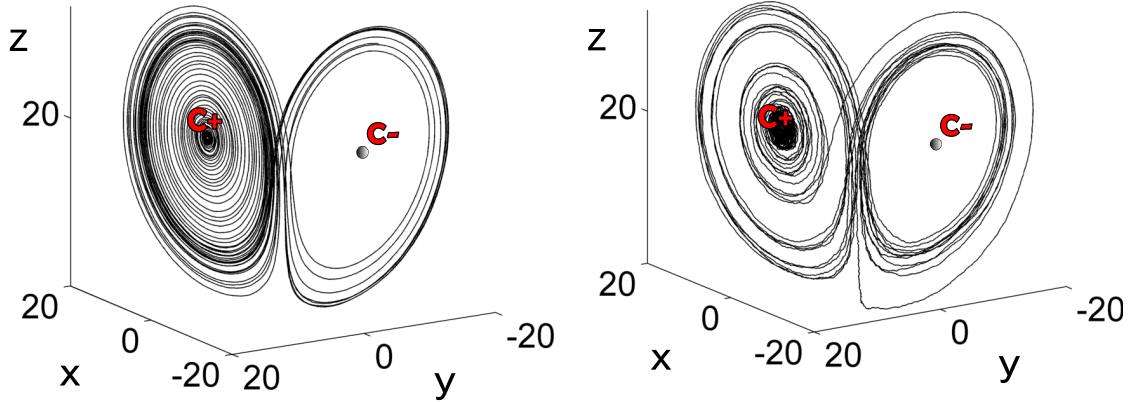}\\
\centering
\caption{\textbf{Dynamics of the Lorenz system.} We select the transient chaotic regime with $\sigma=10$, $b=8/3$ and $r=20$. On the left, the trajectory is deterministic. On the right, the trajectory is affected by some disturbances. The disturbances here, were enlarged in order to help the eye.  Almost all trajectories eventually spiral to one of the two attractors ($C^{+}$ or $C^{-}$). Here both trajectories spiral to $C^{+}$.}
\label{3}
\end{figure}

To describe how the method can be applied to a flow affected by disturbances, we have chosen the Lorenz system \cite{Lorenz}, which is one of the best known models in nonlinear dynamics. This system is a flow, that describes a simplified model of atmospheric convection.  The model consists of three ordinary differential equations,
\begin{eqnarray}
  \dot{x}&=&-\sigma x+\sigma y  \nonumber \\
   \dot{y}&=&-x z + rx-y \\
   \dot{z}&=&x y-b z. \nonumber
\end{eqnarray}
Depending on the parameter values $r$, $\sigma$, and $b$, the system can exhibit different dynamical behaviors, either periodic solutions, chaotic attractors or even transient chaos. Fixing $\sigma=10$, $b=8/3$, transient chaos can be found in the interval $r\in[13.93,24.06]$ as described in \cite{KaplanYorke,YorkeYorke}. For our simulations, we have chosen the value $r=20.0$. In this regime, as we show in the left side of Fig.~\ref{3}, there are transient chaotic orbits that eventually decay towards one of the two point attractors.
\begin{eqnarray}
&C^+=&(\sqrt{b(r-1)},\sqrt{b(r-1)},r-1)\approx(7.12, 7.12, 19)  \nonumber \\
&C^-=&(-\sqrt{b(r-1)},-\sqrt{b(r-1)},r-1)\approx(-7.12, -7.12, 19), \nonumber
\end{eqnarray}
which physically represent a steady rotation of a fluid flow, one clockwise, and the other counterclockwise. Without intervention, transient chaotic trajectories will escape towards these point attractors.

To make this system more realistic, we have added disturbances. The source of the disturbances in a chaotic flow may be diverse, as for example continuous or discrete noise affecting the dynamics, mismatches of the model equations from reality, or the finite precision in the measurement of the state of the system.  This last possibility is especially relevant in chaotic systems since uncertainty grows exponentially with time. In the right side of Fig.~\ref{3}, we show an example of the Lorenz flow affected by some disturbances, where the trajectory eventually spirals to the $C^{+}$ attractor. Our goal here is to apply the partial control method to avoid having trajectories falling to the attractors $C^{+}$ or $C^{-}$ and sustaining them in the transient chaotic regime.

Since the Lorenz system is a flow, different maps can be built, depending on our goals. One important consideration about the application of the method, is how we want to apply the control. One way is to perturb only certain variables of the system. Another possibility is to apply the control only in certain regions of phase space. Alternatively, we can also apply the control at regular times, independently of the state of the system. In all cases, it is important to analyze how the disturbances arise in the map constructed from the flow.

The upper bound of the disturbances in the discrete map could be easily estimated in an experimental setup, measuring the maximum dispersion of an ensemble of trajectories with the same initial condition for a particular Poincar\'e section (2D case) or stroboscopic section (in the 3D case). This procedure could be repeated for several initial conditions taking as upper bound of the disturbances the maximum value of the dispersion found for all the initial conditions tested. This upper bound is the only requirement needed to apply the Sculpting Algorithm to compute the safe set.

\begin{figure}
\includegraphics [trim=0cm 0cm 0cm 0cm, clip=true,width=0.67\textwidth]{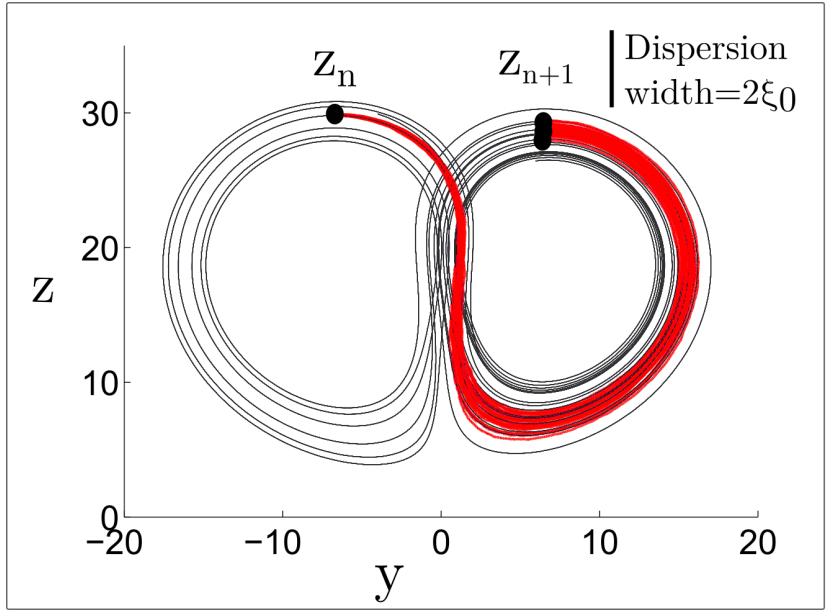}\\
\centering
\caption{\textbf{Possible trajectories. This is the same situation as in Fig.~\ref{3}}. A map constructed taking successive maxima of $z$, with the notation $Z_n$. In red, several trajectories are affected by some small disturbances along the trajectory, all of them starting in the same initial condition. The different trajectories spread out until they reach the next maximum $Z_{n+1}$. Considering the dispersion width in the values of $Z_{n+1}$ as $2\xi_0$, the bound of the effective disturbance affecting the map is $\pm\xi_0$.}
\label{4}
\end{figure}

For example, in Fig.~\ref{4}, we follow Lorenz and discretize the flow by taking the consecutive maxima of the variable $z$, then we obtain a one-dimensional map. We write $Z_n$ for the successive maximum $z$ value. In red, we represent different trajectories affected by different disturbances starting from $Z_n$. As a consequence, the trajectories spread out to yield a dispersion width in $Z_{n+1}$. We can estimate the upper bound of the disturbance $\xi_0$ in the map, as half of the dispersion width.  The dynamics in the map will be $Z_{n+1}=f(Z_n)+ \xi_n$ with $|\xi_n| \leq \xi_0$. For systems of higher dimensions the disturbance in the map can be estimated in the same way. To do that, we have to take every point of the grid and analyze how is the dispersion of the possible trajectories when they return to the map. After that, we take $\xi_0$ as the maximum dispersion observed, recalling that we are assuming bounded disturbances in the dynamics.

Due to the several possibilities for implementing the method in a flow, we describe in the next section three different ways by using a 1D, 2D and 3D map, and discussing the main pros and cons of each choice. We assume in all of them that the upper bound of the disturbances in the map have been previously measured, by using the method described above.

\subsection{1D Map}

\begin{figure}
\includegraphics [trim=0cm 0cm 13cm 0cm, clip=true,width=0.78\textwidth]{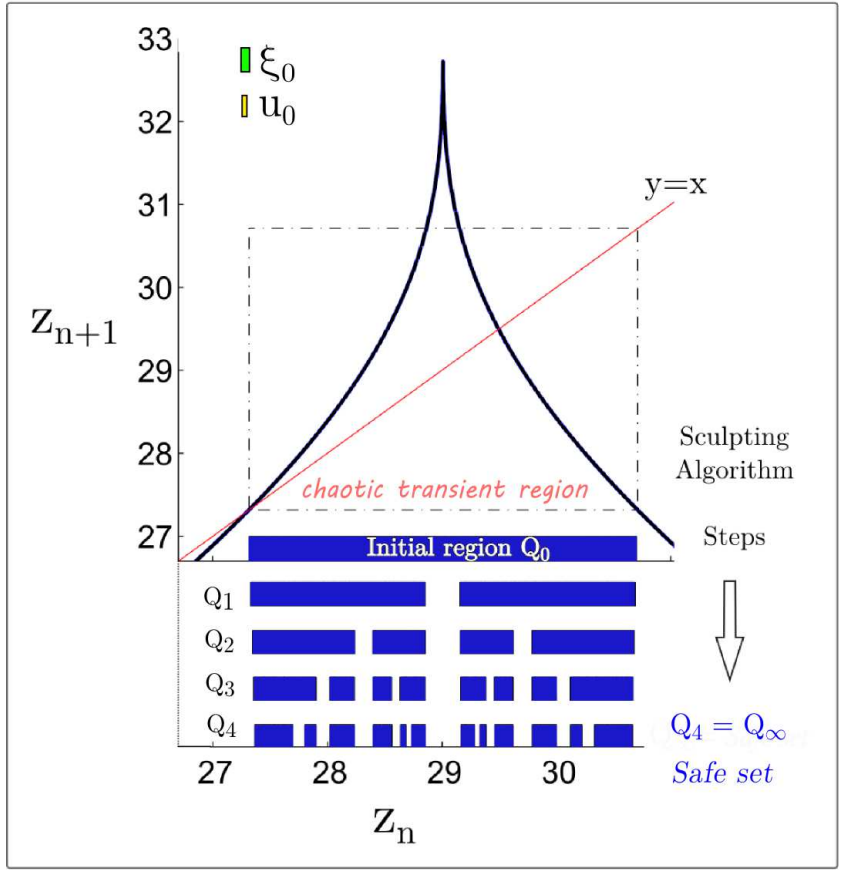}\\
\centering
\caption{\textbf{The 1D safe set.} The black curve is the 1D map built with the successive maxima of $z$. We take as initial set $Q_0$ (upper segment in blue) the region where transient chaos occurs. The map is affected by disturbances with an upper bound $\xi_0=0.080$, while we choose the upper bound of the control as $u_0=0.055$, (the bounds are the width of the bars displayed in the upper left side). The figure shows the successive steps computed by the Sculpting Algorithm, from an initial region $Q_0$ until it converges to the subset $Q_4=Q_\infty \subset Q_0$. We use a grid of $4000$ points in the interval $z_n \in [26.8,30.8]$, that corresponds to a resolution of $0.001$.}
\label{5}
\end{figure}

As shown by Lorenz \cite{Lorenz}, a 1D map for the Lorenz system, can be created by taking the consecutive maxima of the variable $z$. When plotting the pairs $(z_n,z_{n+1})$, one gets (approximately) a function $f$ where $z_{n+1}\approx f(z_n)$. See Fig.~\ref{5}. This is only possible because the sets are very thin. Knowing a local maximum of $z$ is $Z$, allows one to estimate $|x|$ and $|y|$ with considerable precision.

For this map, transient chaos can be observed in the interval $z_n \in [27.3,30.7]$, so we have chosen this interval as the set $Q_0$. We have taken $\xi_0=0.080$. If the control bound $u_0$ is chosen too small, there will be no safe set, and it will be impossible to prevent escapes. In this case, we have taken as the control bound $u_0=0.055$ ($u_0<\xi_0$) which is approximately the minimum value for which a safe set exists. Then, we have obtained the safe set by using the recursive Sculpting Algorithm. In Fig.~\ref{5}, we can see how the algorithm sculpts the initial region $Q_0$ until it finds $Q_4$ where it converges, so $Q_4=Q_\infty$ is the safe set. For this computation we have used a grid of $4000$ points in the interval $z_n \in [26.8,30.8]$, so the grid resolution is $0.001$.

\begin{figure}
\includegraphics [trim=0cm 0cm 0cm 0cm, clip=true,width=0.85\textwidth]{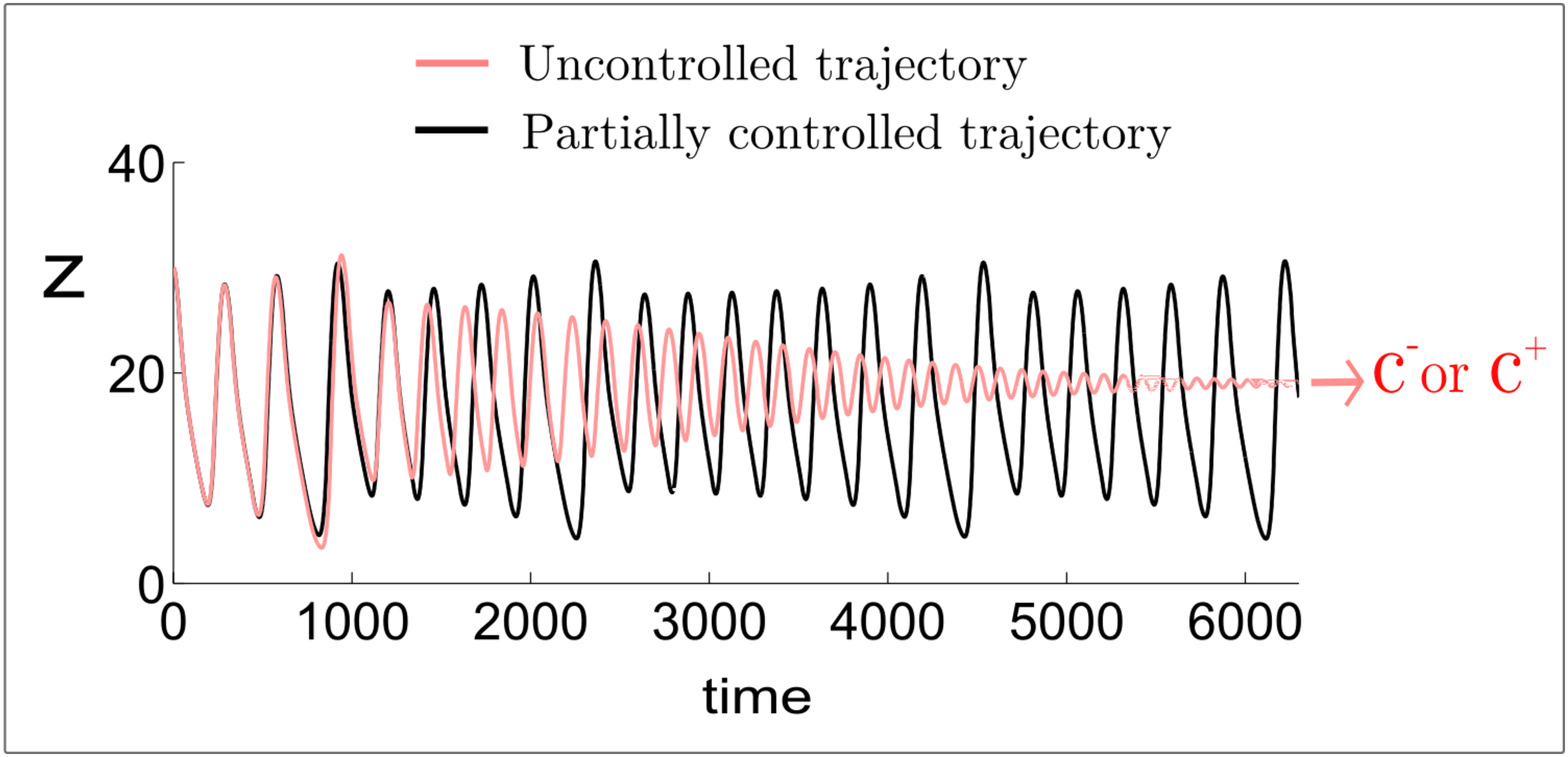}\\
\centering
\caption{\textbf{Time series of the variable $z$ for the Lorenz system with $r=20$.} The figure shows a comparison between an uncontrolled trajectory that escapes from chaos (red line) and a partially controlled trajectory (black line). Starting with the same initial condition, the uncontrolled trajectory eventually decays to $C^{+}$ or $C^{-}$, which physically means a steady rotation of the fluid flow. On the other hand the partially controlled trajectory is maintained in the chaotic transient regime, that is, the rotation of the fluid flow remains chaotic forever.}
\label{6}
\end{figure}

The safe set computed ensures for any starting point $q_n$ in the safe set and any allowable $\xi_n$, there is a $u_n$ that puts $f(q_n)+\xi_n+u_n$ back in the safe set. This is true for the map, however, the control is applied in the phase space so we must take into account of the fact that as each local maximum of $z$ is described by $3$ coordinates $(x_m,y_m,z_m)$, the total distance to the safe set is is $d=\sqrt{(x_m-x_{msafe})^{2}+(y_m-y_{msafe})^{2}+(z_m-z_{msafe})^{2}}$, where $(x_{msafe},y_{msafe},z_{msafe})$ is the closest point belonging to the safe set. In Fig.~\ref{6} we show a controlled time series of the $z$ variable in contrast with an uncontrolled trajectory. We can see that chaos is sustained by applying small perturbations in the maxima of the variable $z$.

The main advantage of this 1D approach is that the computation of the safe set is easy and fast. This kind of map is useful when the disturbed trajectories mainly spread out along the expanding direction of the chaotic saddle, as occurs in the case of stochastic noise or uncertainties in the application of the controls. See for example in Ref.~\cite{Ecology}, where an ecological model of three species was studied. In that case, it was possible to construct a map of the form $(x_i,y_i,z_i)$ where $y$ and $z$ kept constant, and only $x$ changed after one iteration of the map. That kind of situations allows the control of system while  perturbing only one of the variables.

\subsection{2D Map}

\begin{figure}
\includegraphics [trim=0cm 0cm 0cm 0cm, clip=true,width=0.65\textwidth]{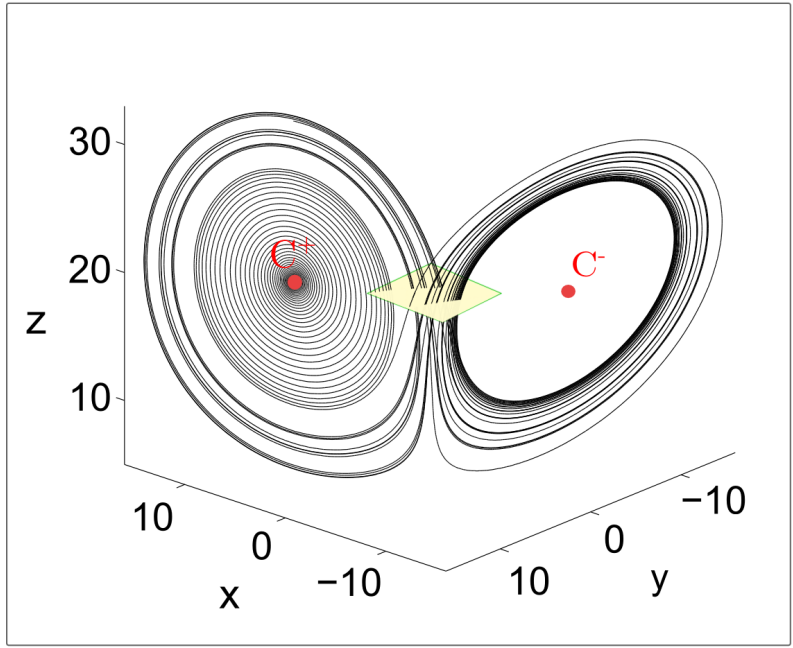}\\
\centering
\caption{\textbf{The Lorenz system with $r=20$ (transient chaos).} The figure shows an uncontrolled trajectory in phase space crossing a square with $x\in[-3,3]$ and $y\in[-3,3]$ in the plane $z=19$. To built the map, we use a grid of initial conditions in the plane, and evaluate the images of the trajectories when they cross again the plane. The goal of the control will be to keep the trajectories in this plane, avoiding the escape to one of the attractors $C^+$ or $C^-$, placed outside.}
\label{7}
\end{figure}

In the case of three-dimensional flows, one can build a discrete 2D map taking a Poincar\'{e} or surface section that intersects the flow. For our purpose, we have chosen the plane $z=19$ with the ranges $x\in[-3,3]$ and $y\in[-3,3]$, as shown in Fig.~\ref{7}. The trajectories that cross this plane are in the transient chaotic regime, while the attractors $C^+=(7.12, 7.12, 19)$ and $C^-=(-7.12, -7.12, 19)$ that we want to avoid, are situated outside this plane (see the location in Fig.~\ref{7}). For this reason, we have taken as $Q=Q_0$, the square $x\in[-3,3]$ and $y\in[-3,3]$ in the plane $z=19$. Then we have used the Sculpting Algorithm to find the safe set $Q_\infty \subset Q$, designed to avoid the eventually decay to the attractors.

\begin{figure}
\includegraphics [trim=0cm 0cm 0cm 0cm, clip=true,width=0.6\textwidth]{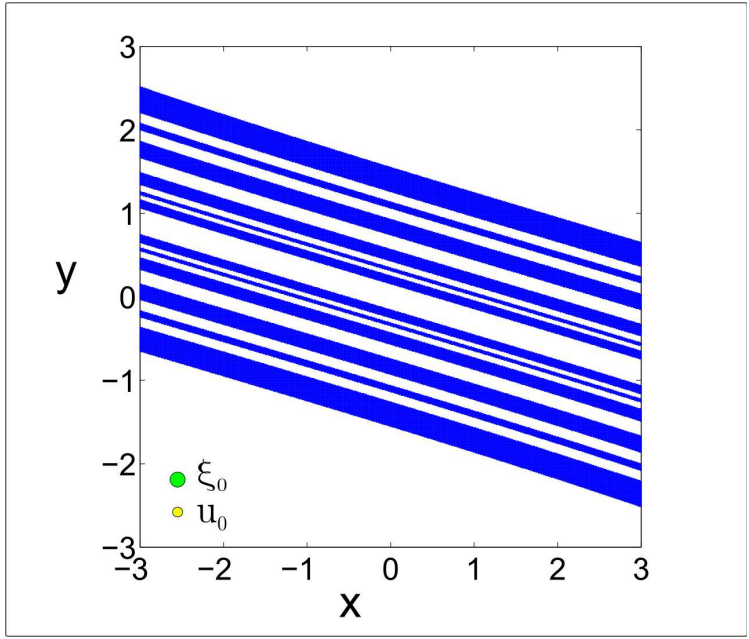}\\
\centering
\caption{\textbf{The 2D safe set.} The safe set obtained using the map built with the plane displayed in Fig.~\ref{7}. We show in blue the computed safe set $Q_\infty$ for $\xi_0=0.09$ and $u_0=0.06$ ($u_0<\xi_0$). The grid size used is $1201 \times 1201$ points. The radius of the balls in the lower left side indicates the bounds of the disturbance, $\xi_0=0.09$ (green) and the control $u_0=0.06$ (yellow).}
\label{8}
\end{figure}

\begin{figure}
\includegraphics [trim=0cm 0cm 0cm 0cm, clip=true,width=0.65\textwidth]{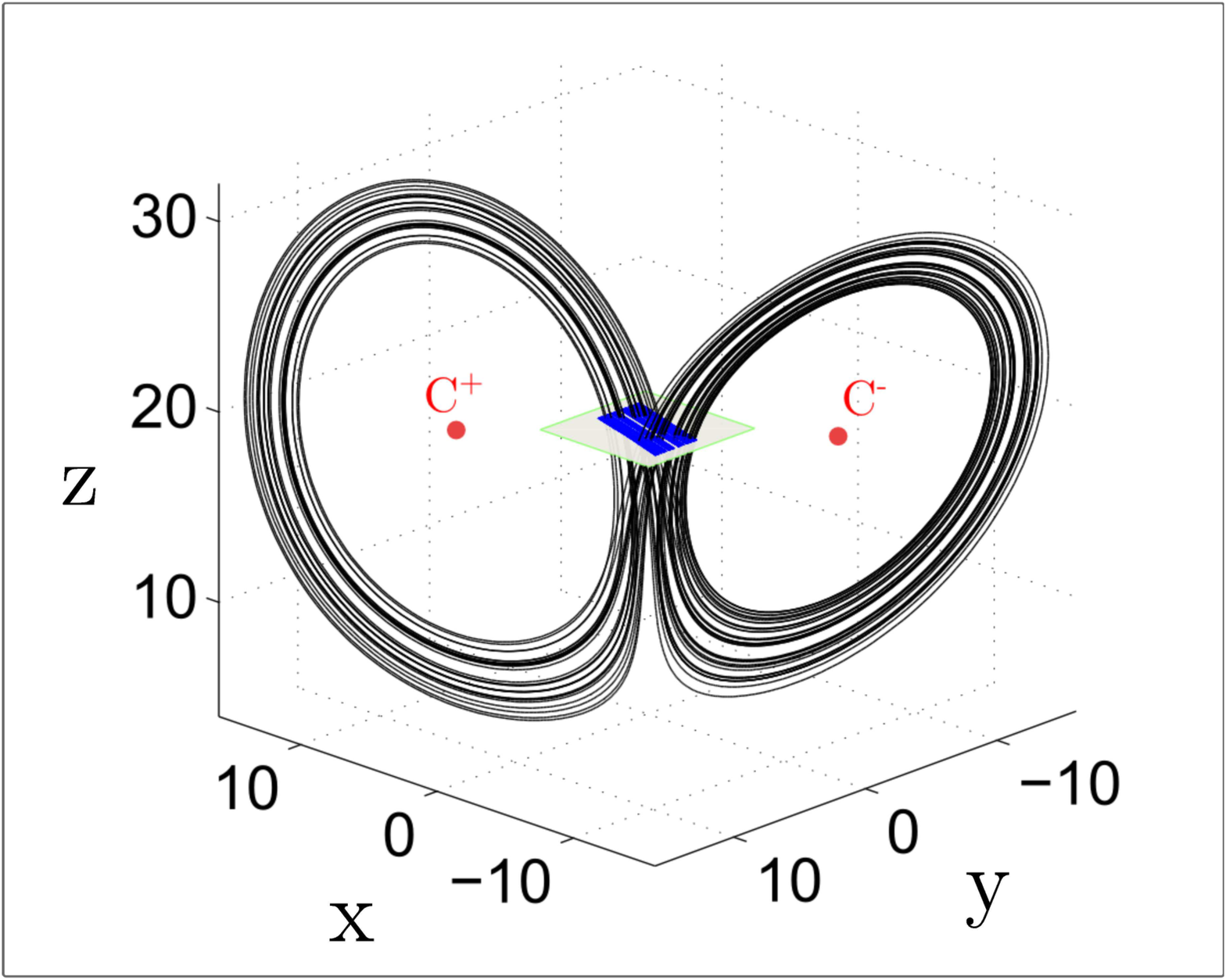}\\
\centering
\caption{\textbf{A partially controlled trajectory.} Here we see a partially controlled trajectory in phase space for case in Fig.~\ref{8}. Each time that the trajectory crosses the safe set plane (placed in $z=19$), the control is applied pushing the trajectory onto the set in Fig.~\ref{8} avoiding the escape from chaos. In addition, the partially controlled trajectory remains chaotic.}
\label{9}
\end{figure}

\begin{figure}
\includegraphics [trim=0cm 0cm 0cm 0cm, clip=true,width=0.56\textwidth]{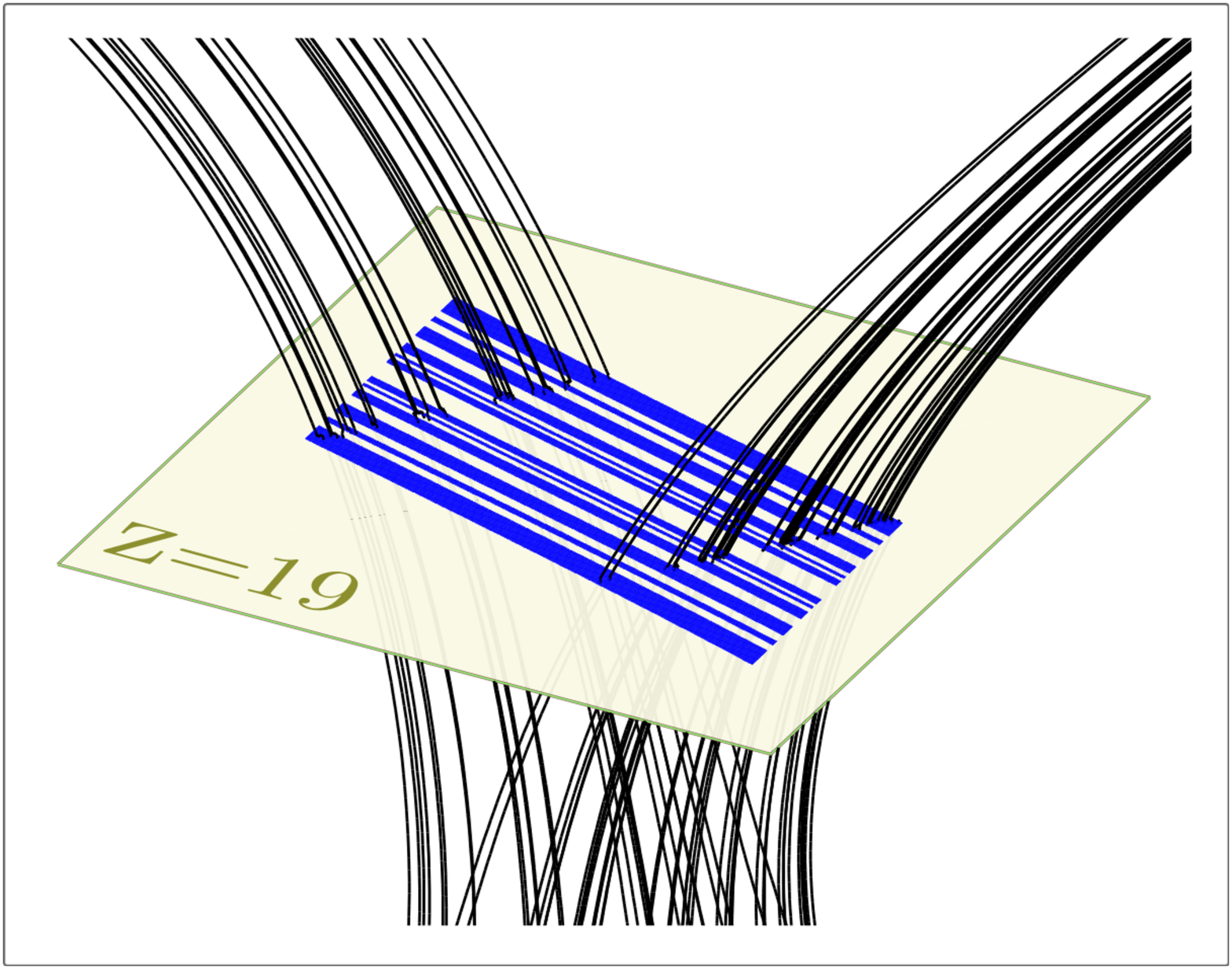}\\
\centering
\caption{\textbf{Detailed viewed of a partially controlled trajectory}. Here the situation is the same as in Figs.~\ref{8} and ~\ref{9}. The figure shows a zoom of the Fig.~\ref{9}, centered in the safe set where the control is applied. Small controls $u_n$ are applied when crossing the plane $z=19$ to force the trajectories (in black) to pass through the safe set (in blue).}
\label{10}
\end{figure}

 As an example, we have assumed that the map is affected by some disturbances with upper bound $\xi_0=0.09$. Applying the Sculpting Algorithm, we have found the safe set for the minimum possible value of the control, that is $u_0=0.06$ ($u_0<\xi_0$). In Fig.~\ref{8}, the resultant safe set is displayed. A partially controlled trajectory is represented in Fig.~\ref{9}, where we have also shown the safe set in phase space in order to see how it is used to control the system. Notice that, we are able to avoid the attractors, applying only small perturbations in the plane. A zoom of this region is shown in Fig.~\ref{10}. The computation was carried out taking a grid size of $1201 \times 1201$ points, (grid resolution is $0.005$ in both variables $x$ and $y$).

When a map comes from a Poincar\'{e} cross section, one can deals with systems where all the variables are affected by some bounded disturbance. In addition, as opposed to the 1D map, where we have to act on the $x$, $y$ and $z$ variables to control the system, the control in the 2D map is only applied in the variables $x$ and $y$, since $z$ is constant. This can be an advantage in systems where it is difficult or expensive to apply the control in each variable.

\subsection{3D Map}

The 1D approach as well as the 2D approach, have the disadvantage of having to track the trajectory to know when it passes through the control region, where we apply the control corrections. Another strategy is to put the focus on the time instead of the variables. In this way, it is possible to apply the partial control method using a time discretization of the Lorenz system, which allows us to obtain a 3D discrete map. With this kind of map it is possible to avoid the escapes from chaos by applying the control with a fixed time interval, which can be a useful practice in many situations. The 3D map is obtained from the 3D flow by taking a suitable time interval $\Delta t$ between the current state of the system and the future state, that is, $x(t_0), y(t_0), z(t_0) \rightarrow x(t_0+\Delta t),y(t_0+\Delta t),z(t_0+\Delta t)$. By computing the time-$\Delta t$ image of each point of a 3D grid that cover the phase space, we can obtain the 3D map.

The choice of $\Delta t$ is important, since it is related with the topology of the map obtained. If $\Delta t$ is too small no safe sets exist (given $u_0 < \xi_0$), while for a sufficiently large time interval, the safe set appears. The topological explanation for this, is that the flow is acting like a pastry transformation which takes some time to be completed. Once this time is reached, the safe set appears. For our Lorenz system, there are safe sets for values of $\Delta t \geq 1.2$.

\begin{figure}
\includegraphics [trim=0cm 0cm 0cm 0cm, clip=true,width=1\textwidth]{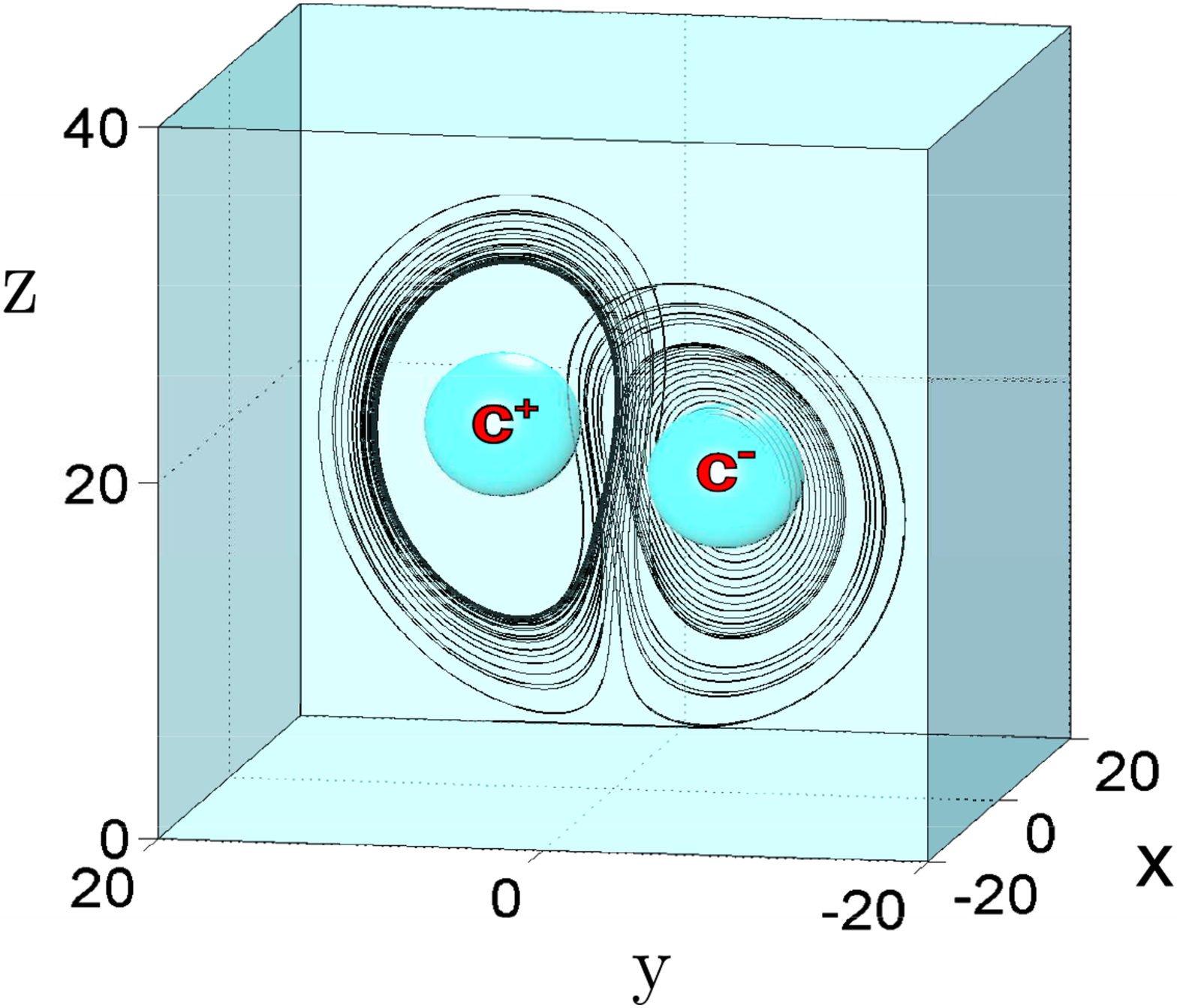}\\
\centering
\caption{\textbf{A choice of 3D set $Q$}. The 3D set $Q$ is the cube $x\in[-20,20]$, $y\in[-20,20]$, $z\in[0,40]$ except that the balls of radius $4$, centered in $C^+=(7.12, 7.12, 19)$ and $C^-=(-7.12, -7.12, 19)$ are removed from $Q$. We want trajectories to stay in $Q$ and not fall to these attractors. A trajectory is plotted to show the chaotic transient behavior in this region.}
\label{11}
\end{figure}

\begin{figure}
\includegraphics [trim=0cm 0cm 0cm 0cm, clip=true,width=0.62\textwidth]{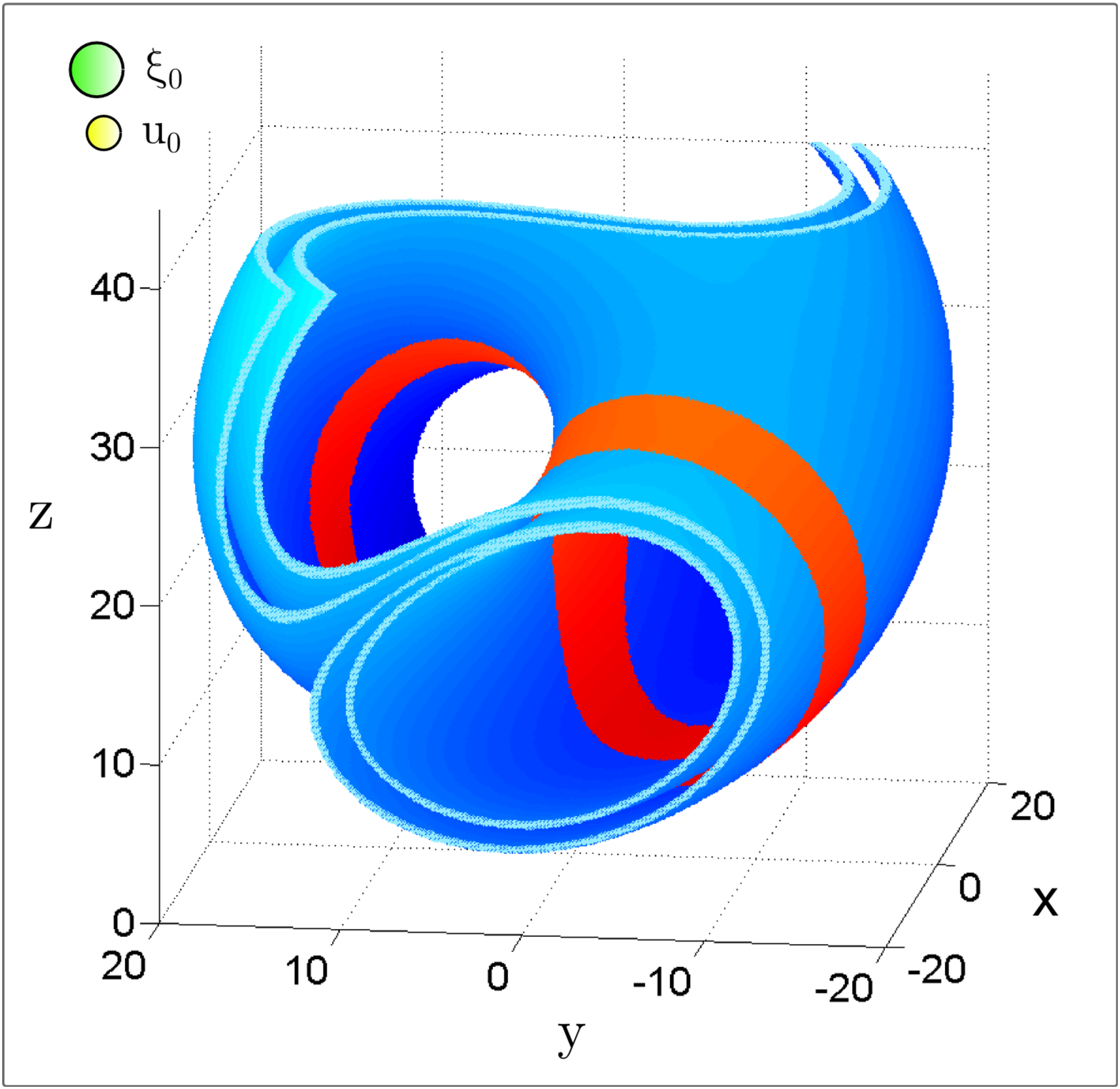}\\
\centering
\caption{\textbf{The 3D safe set.} We show in blue the 3D \emph{safe set} $Q_\infty$ for Fig.~\ref{11}, obtained after applying the Sculpting Algorithm. We set $\Delta t=1.2$, $\xi_0=1.5$ ($\xi_0=$ radius of the green ball) and $u_0=1.0$ ($u_0=$ yellow ball's radius). We show in red the asymptotic safe set which is a subset of the safe set. This is the region in which the controlled trajectories eventually lie.}
\label{12}
\end{figure}

For a 3D example, we take the domain with $x\in[-20,20]$, $y\in[-20,20]$, $z\in[0,40]$, with a grid size of $401 \times 401 \times 401$, so the grid resolution is $0.1$ for each variable. In this region, there are transient chaotic trajectories, which eventually decay to the attractors $C^+=(7.12, 7.12, 19)$ and $C^-=(-7.12, -7.12, 19)$.  As we want to avoid  $C^+$ and $C^-$, balls centered in these attractors are removed. See the region $Q$ and a transient chaotic trajectory in Fig.~\ref{11}. To obtain the map, we have computed the image of each point of $Q$ with $\Delta t=1.2$. Then, as an example, we take the value $\xi_0=1.5$ and $u_0=1.0$  (note $u_0 < \xi_0$). Using the Sculpting Algorithm, we obtain the safe set shown in Fig.~\ref{12}.

To describe the controlled dynamics in the 3D map we write $q_n$ for the controlled trajectory at time $n \Delta t$. To obtain a particular trajectory we choose $\xi_n$ at random with $|\xi_n| \leq \xi_0$. Then we choose some $u_n$, that place  $q_{n+1}=q_n +\xi_n + u_n$ in the safe set. In each case, $\xi_n$ represents the disturbance accumulated by the trajectory in the time interval $\Delta t$, while the control is always applied at a discrete time. Notice that the requirement $|u_n| \leq u_0$ allows for a flexible control, since for most iterations there is more than one point belonging to the safe set which can be reached without exceeding the upper control bound $u_0$.  In this case, we apply the minimum control, which is almost always unique.

\begin{figure}
\includegraphics [trim=0cm 0cm 0cm 0cm, clip=true,width=0.69\textwidth]{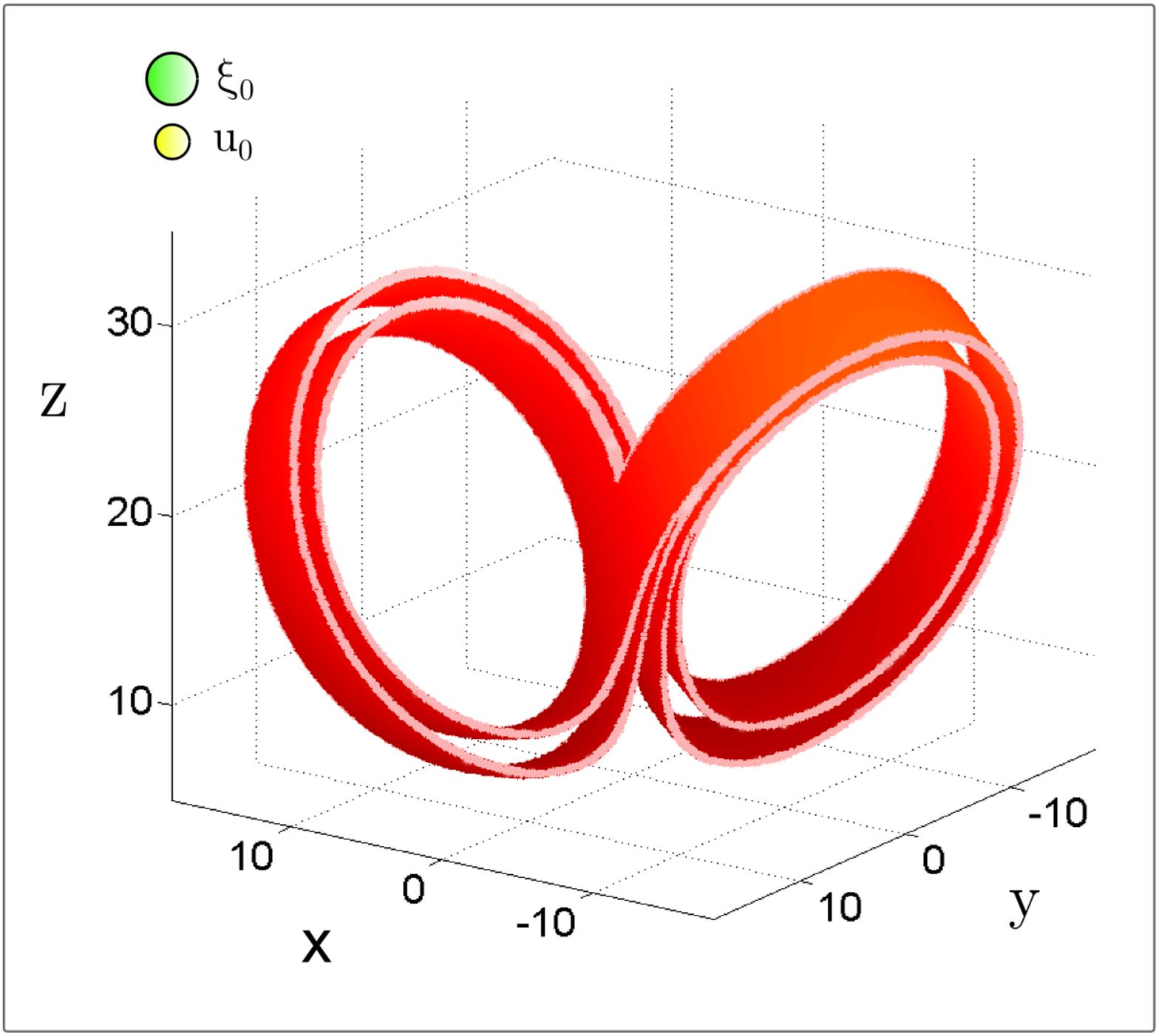}\\
\centering
\caption{\textbf{The asymptotic safe set.} The situation as in Fig.~\ref{12}. We show only the asymptotic safe set. Partially controlled trajectories converge rapidly to this region.}
\label{13}
\end{figure}

\begin{figure}
\includegraphics [trim=0cm 0cm 0cm 0cm, clip=true,width=0.67\textwidth]{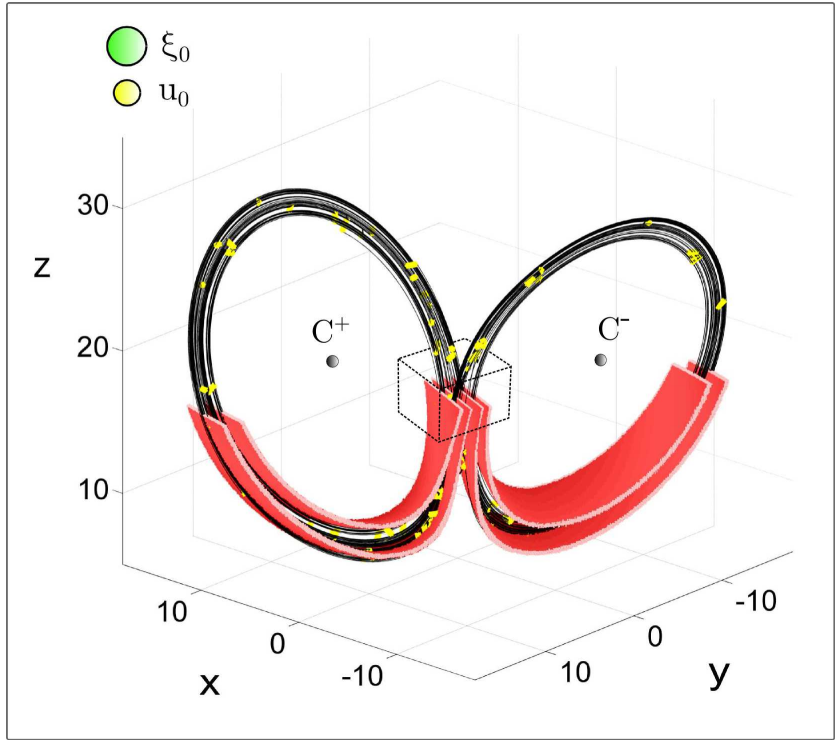}\\
\centering
\caption{\textbf{Asymptotic safe set with a partially controlled trajectory.} The situation is the same as in Figs.~\ref{12} and \ref{13}. Here we display a cut-away section of the asymptotic safe set in order to see a partially controlled trajectory (with $\Delta t=1.2$) displayed in black.  The controls (yellow segments inserted in the trajectory) are applied every $\Delta t=1.2$. As a result, the trajectory is kept in the chaotic region and the attractors $C^+$ and $C^-$ are avoided. }
\label{14}
\end{figure}

\begin{figure}
\includegraphics [trim=0cm 0cm 0cm 0cm, clip=true,width=0.46\textwidth]{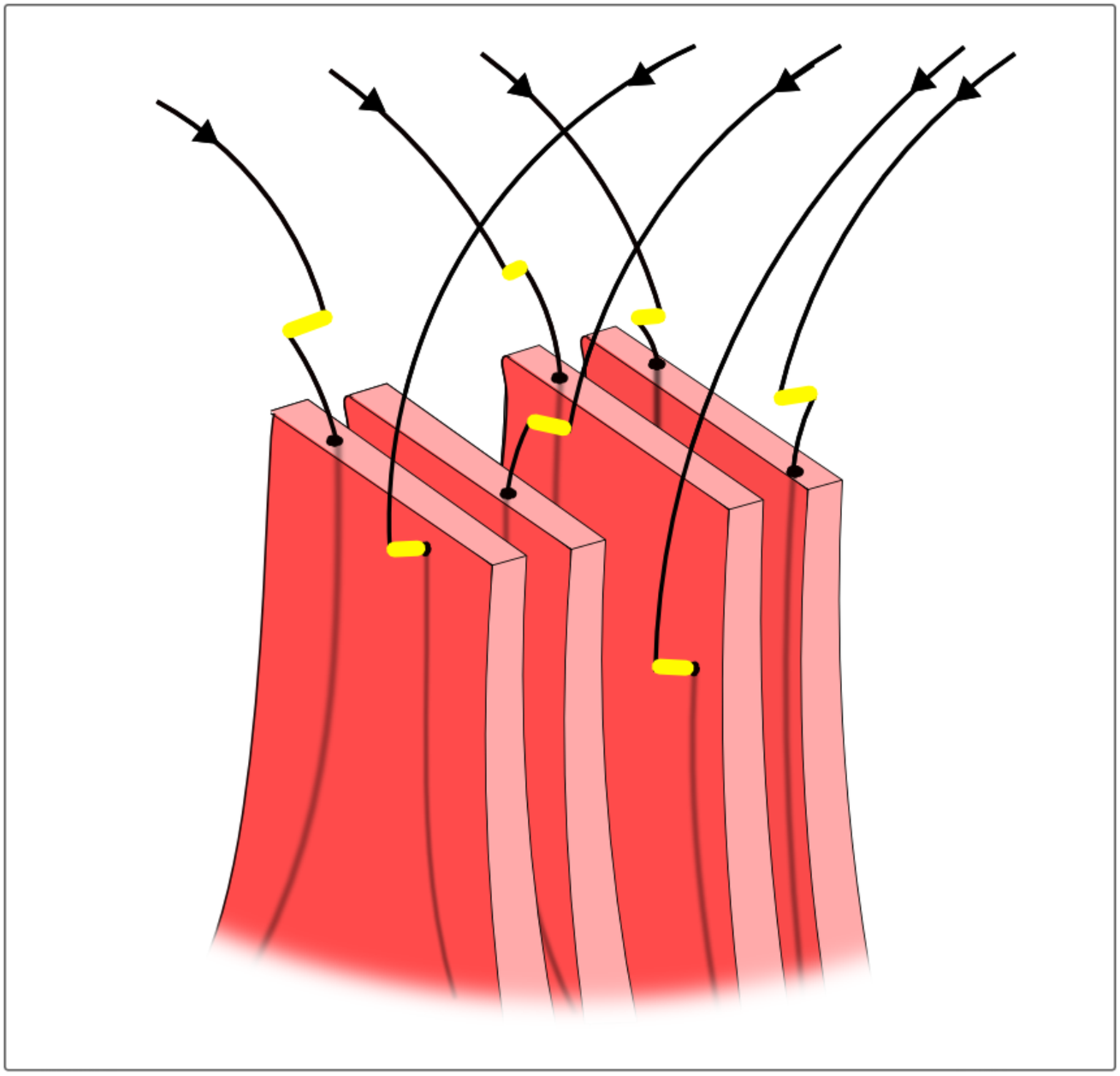}\\
\centering
\caption{\textbf{A detailed view of Fig.~\ref{14}.} The figure is a zoom in on the small cube displayed in Fig.~\ref{14}. Only few lines are displayed for a better visualization. The controls (yellow segments) are applied to move the trajectories (in black) into the asymptotic safe set (in red).}
\label{15}
\end{figure}

One interesting feature of the partial control method is that the controlled trajectories converge towards a certain region of the safe set, which is called the \textbf{asymptotic safe set}, (see Figs.~\ref{12} and~\ref{13}). Controlled trajectories do not leave the asymptotic safe set once they reach it, (unless the control is turned off). Once the dynamics converges, it is sufficient to use the asymptotic safe set to control the trajectories. In Fig.~\ref{14}, a partially controlled trajectory is displayed. The controls, represented as yellow segments distributed along the trajectory, are applied every $\Delta t=1.2$. We show this fact with a zoom in Fig.~\ref{15}. As a result, the trajectories never fall into the attractors $C^+$ or  $C^-$, keeping the dynamics in the chaotic region forever.

\begin{figure}
\includegraphics [trim=0cm 0cm 0cm 0cm, clip=true,width=0.68\textwidth]{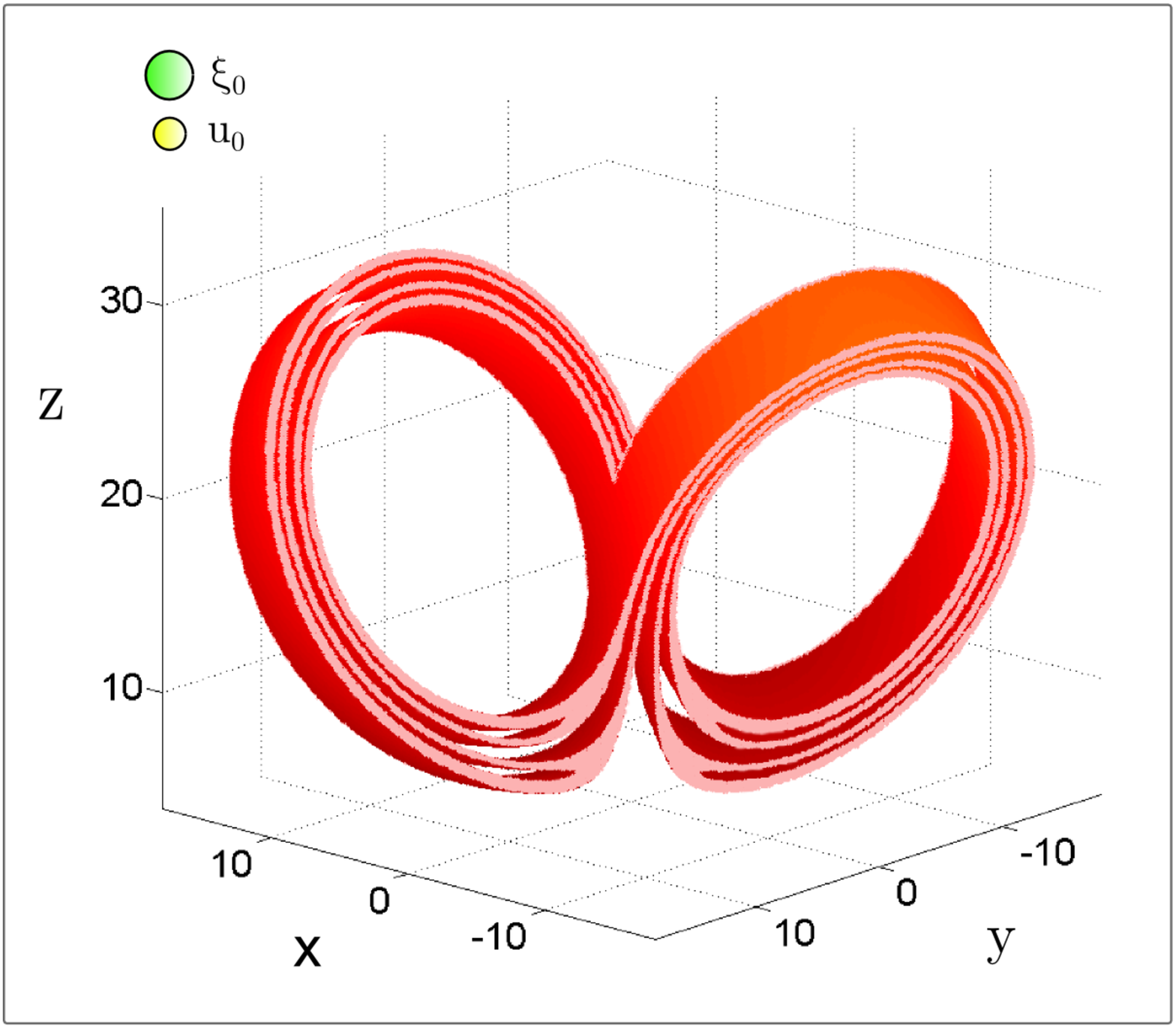}\\
\centering
\caption{\textbf{The asymptotic safe set computed for $\Delta t=1.8 $.} To compute this set we have taken $\xi_0=1.5$ (green ball) and $u_0=1.0$ (yellow ball).}
\label{16}
\end{figure}

\begin{figure}
\includegraphics [trim=0cm 0cm 0cm 0cm, clip=true,width=0.68\textwidth]{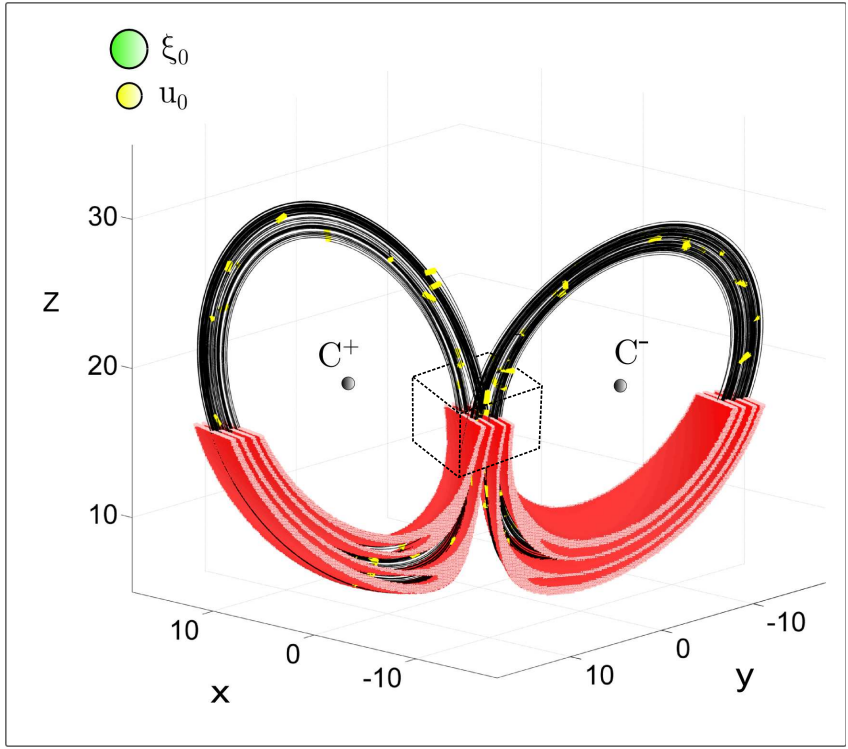}\\
\centering
\caption{\textbf{Partially controlled trajectories in the asymptotic safe set for $\Delta t=1.8 $.} Same situation as in Fig.~\ref{16}. The figure displays a half section of the asymptotic safe set in order to visualize a partially controlled trajectory (in black). In this case the controls (yellow segments inserted in the trajectory) are applied every $\Delta t=1.8$ instead of $\Delta t=1.2$ as in the previous example. The zoom of the small cube in the center, has a similar appearance as the zoom displayed in Fig.~\ref{14}. The resulting  partially controlled trajectory is kept in the chaotic region and the attractors $C^+$ and $C^-$ are avoided.}
\label{17}
\end{figure}

As we have mentioned, the safe set appears for values of $\Delta t \geq 1.2$, so it is possible to adapt the control frequency to our specific requirements, taking other $\Delta t$ values. Figure~\ref{16} shows the asymptotic safe set for $\Delta t=1.8$ , and with $\xi_0$ and $u_0$ unchanged. With this set we could control the system applying a control every $\Delta t=1.8 $ (see Fig.~\ref{17}) instead of $\Delta t=1.2 $ as in the previous case. Which choice of $\Delta t$, $1.2$ or $1.8$ is better to minimize the control?. It depends on how the disturbances affect the trajectories. For example, it is common in most scenarios that the cumulative effect of disturbances grows exponentially with time due to chaos, and therefore the needed $u_0$ increases as well \cite{Frequency}, so it is a question of balance between the suitable time interval and the disturbance arising in the map.


\begin{figure}
\includegraphics [trim=0cm 0cm 0cm 0cm, clip=true,width=0.8\textwidth]{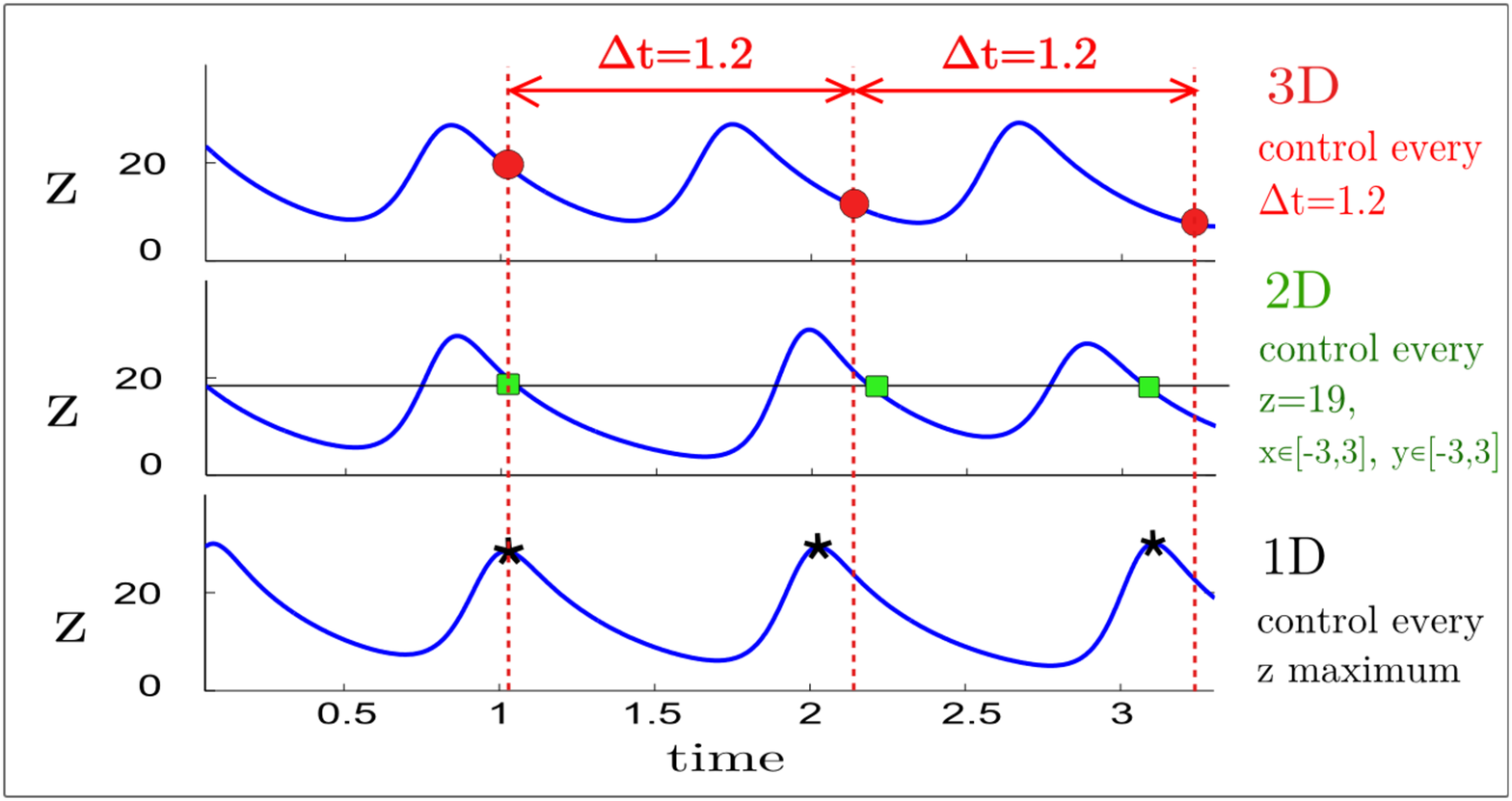}\\
\centering
\caption{\textbf{Comparison of the three controlled trajectories of the $z$ variable obtained with the 3D, 2D and 1D map respectively.} The marks indicate the points where the control is applied. Only in the 3D case are the controls time periodic.}
\label{18}
\end{figure}

Using a discretization with fixed $\Delta t$ time intervals can be advantageous. This strategy can provide a possibly useful way to control real situations. For example, in the context of medicine, a medical treatment based on the partial control method, could be applied a fixed day of the week, which supposes an easy and convenient control relationship between the physician and the patient. To highlight this feature, we compare in Fig.~\ref{18}, three controlled trajectories obtained with the respective map (3D, 2D and 1D). We have marked in the figure, the points where the control is applied. Notice that, unlike the other maps, in the 3D map it is possible to apply the control with a fixed time interval.

\section{Conclusions}

For the application of the partial control method a very few amount of ingredients are required. We only need a transient chaotic system with escapes, the knowledge of the upper bound of the disturbances and an upper bound control high enough to find a safe set with the Sculpting Algorithm. We believe that these conditions are rather general. In the real time application of the control, the controller only needs to know which is the state of the system and which is the safe set. If the state of the system is in the safe set no control is applied, whereas if the state of the system is not in the safe set, a small amount of control is needed to put the system inside the safe set again.

We have applied the partial control method to the Lorenz system in the presence of disturbances, for a particular choice of parameters where it shows transient chaos. Typical uncontrolled trajectories in this system follow a chaotic motion until they escape to one of its two stable non-chaotic attractors. With the goal of avoiding these escapes, we have applied the partial control method in three different ways. We have built 1D, 2D and 3D maps, and obtained the respective safe sets with the Sculpting Algorithm.

Using the respective safe sets in each case, we have shown that is possible to control the trajectories, using a small amount of control in comparison with the disturbances affecting the system. Another remarkable feature is that the partially controlled trajectories keep the chaotic behavior of the original system. Since $u_0< \xi_0$, it is impossible for the controller to completely determine the oscillatory behavior.

The possibility of using different kinds of maps to control the dynamics allows us flexibility. However, in some situations it can be convenient to apply the control in periodic time intervals. This strategy is shown in the 3D case with a fixed time discretization $\Delta t$. This novel approach, allows us to focus the attention only in the time instead of the control region. In addition, the frequency of these controls can be adapted depending on the specific experimental requirements, which can suppose an easy and flexible way to control the system.

Finally, we want to highlight the potential of this control approach. The Lorenz system was considered here, because is a very well known and paradigmatic system in nonlinear dynamics and it shows in a clear way how the partial control method works. Many other dynamical systems that show transient chaos with undesirable escapes can be controlled using a similar procedure.

\enlargethispage{20pt}

\begin{acknowledgments}
Financial support from the Spanish Ministry of Economy and Competitiveness under Project No. FIS2013-40653-P is acknowledged.
\end{acknowledgments}


\begin{thebibliography}{9}

\bibitem{ControlD}
Bradley E, Zhao F.
 Phase-Space Control System Design.
 {\em IEEE Control Syst. Mag.}, vol. 13, no. 2, pp. 39-46, 1993.

\bibitem{ControlCha}
Vincent TL.
  Control Using chaos.
 {\em IEEE Control Syst. Mag.}, vol. 17, no. 6, pp. 65-76, 1997.

\bibitem{StabilityDelay}
Sipahi R, Niculescu SI, Abdallah CT, Michiels M.
 Stability and Stabilization of Systems with Time Delay,''
 {\em IEEE Control Syst. Mag.}, vol. 31, no. 1, pp. 38-65, 2011.


\bibitem{Oscillator}
Schwartz IB, Triandaf I.
 The slow invariant manifold of a conservative pendulum-oscillator system.
 {\em International Journal of Bifurcation and Chaos} 1996; 6:673-692.

\bibitem{Thermal}
In V, Spano ML, Neff JD, Ditto WL, Daw CS, Edwards KD, Nguyen K.
 Maintenance of chaos in a computational model of thermal pulse combustor.
 {\em Chaos} 1997; 7:605-613.

\bibitem{Perc}
Perc M, Marhl M.
 Chaos in temporarily destabilized regular systems with the slow passage effect.
{\em Chaos, Soliton and Fractals.} 2006; 27:395-403.

\bibitem{Biological}
Yang W, Ding M, Mandell AJ, Ott E.
 Preserving chaos: Control strategies to preserve complex dynamics with potential relevance to biological disorders.
 {\em Physical Review E} 1995; 51:102-110.

\bibitem{Laser}
Dangoisse D, Glorieux P, Hannequin D.
 Laser chaotic attractors in crisis.
 {\em Phys. Rev. Lett.} 1986; 57:2657-2660.

\bibitem{Dhamala}
Dhamala M, Lai YC.
 Controlling transient chaos in deterministic flows with applications to electrical power systems and ecology.
 {\em Phys. Rev. E} 1999; 59:1646-1655.


\bibitem{McCann}
McCann K, Yodzis P.
 Bifurcation structure of a three-species food chain model.
 {\em Theor. Popul. Biol.} 1995; 48:93-125.


\bibitem{Ecology}
Cape\'ans R, Sabuco J, Sanju\'an MAF.
 When less is more: Partial control to avoid extinction of predators in an ecological model.
 {\em Ecol. Complex} 2014; 19:1-8.

\bibitem{Cancer}
Lop\'ez AG, Sabuco J, Seoane JM, Duarte J, Janu\'ario C, Sanju\'an MAF.
 Avoiding healthy cells extinction in a cancer model.
 {\em J. Theor. Biol.} 2014; 349:74-81.

\bibitem{Schwartz}
Schwartz IB, Triandaf I.
 Sustainning chaos by using basin boundary saddles.
 {\em Phys. Rev. Lett.} 1996; 77:4740-4743.


\bibitem{Bertsekas}
Bertsekas DP.
Infinite-time reachability of state-space regions by using feedback control.
{\em IEEE Trans. Autom. Control}, vol. 17, no. 5, pp. 604-613, 1972.

\bibitem{Bertsekasdos}
Bertsekas DP and Rhodes IB.
On the minimax reachability of target set and target tubes.
 {\em Automatica}, vol. 7, pp. 233-247, 1971.

\bibitem{Automatic}
Sabuco J, Zambrano S, Sanju\'an MAF, Yorke JA.
 Finding safety in partially controllable chaotic systems.
 {\em Commun. Nonlinear Sci. Numer. Simul.} 2012; 17:4274-4280.


\bibitem{Asymptotic}
Sabuco J, Zambrano S, Sanju\'an MAF, Yorke JA.
 Dynamics of partial control.
 {\em Chaos} 2012; 22,047507.


\bibitem{Invariant}
Blanchini F.
 Set invariance in control.
 {\em Automatica} 1999; 35(11):1747-1767.

\bibitem{Hutson}
Hutson V, Schmitt K.
 Permanence and the dynamics of biological systems.
 {\em Math. Biosci.} 1992; 111:1-71.


\bibitem{Genesio}
Genesio R, Tartaglia M, and Vicino A.
On the estimate of asymptotic stability regions: Stateof art and new proposal.
{\em IEEE Trans. Autom. Control}, vol. 30, no. 8, pp. 747-755, 1985.


\bibitem{Kolmanovski}
Kolmanovski IV and Gilbert EG.
 \textit{Multimode regulators for systems with state and control constraints and disturbance inputs.}
 Berlin:Springer, 1997.


\bibitem{Gutmandos}
Gutman PO and Cwikel M.
Convergence of an algorithm to find maximal state constraint sets for discrete-time linear dynamical systems with bounded control and states.
 {\em IEEE Trans. Autom. Control}, vol. 31, no. 5, pp. 457-459, 1986.

\bibitem{Gutman}
Gutman PO, Cwikel M.
 Admisible sets and feedback control for discrete-time linear systems with bounded control and states.
 {\em IEEE Trans. Autom. Control} 1986; 31(4):373-376.


\bibitem{Gutmantres}
Gutman PO and Cwikel M.
An algorithm to find maximal state constraint sets for discretetime linear dynamical systems with bounded control and states.
 {\em IEEE Trans. Autom. Control}, vol. 32, no. 3, pp. 251-254, 1987.

\bibitem{Astrom}
Astr\"{o}m KJ.
 \textit{Event based control.}
Berlin:Springer, 2008.


\bibitem{Donkers}
Donkers MCF, Tabuada P, and Heemels WPMH.
 Minimum attention control for linear systems.
 {\em  Discret Event Dyn. S.}, vol. 24, no. 2, pp. 99-218, 2012.

\bibitem{Nagumo}
Nagumo M.
\"{U}ber die Lage der Integralkurven gew\"{o}hnlicher Differentialgleichungen.
 {\em Proc. Phys.-Math. Soc. Japan}, vol. 24, no. 3, pp. 551-559, 1942.

\bibitem{Das}
Das S and Yorke JA.
Avoiding extremes using partial control.
 {\em J. Differ. Equations Appl.}, 2015.


\bibitem{Lorenz}
Lorenz E.
 Deterministic nonperiodic flow.
 {\em J. Atmos. Sci.} 1963; 20:130-141.

\bibitem{KaplanYorke}
Kaplan JL, Yorke JA.
 Preturbulence: A regime observed in a fluid flow model of Lorenz.
 {\em Commun. Math. Phys.} 1979; 67:93-108.

\bibitem{YorkeYorke}
Yorke JA and Yorke ED.
 Metastable chaos: the transition to sustained chaotic behavior in the Lorenz model.
 {\em J. Stat. Phys.} 1979; 21:263-277.

\bibitem{Brain}
Schiff SJ, Jerger K, Duong DH, Chang T, Spano ML and Ditto WL.
 Controlling chaos in the brain.
{\em Nature.}, vol. 370, pp. 615-620, 1994.

\bibitem{ExpChaos}
Ditto WL, Rauseo SN and Spano ML.
 Experimental control of chaos.
 {\em Phys. Rev. Lett}, vol. 65, pp. 3211, 1990.

\bibitem{codes}
 See supplementary material at http://www.fisica.urjc.es/physics/partialcontrol for the software for partial control.

\bibitem{Frequency}
Zambrano S, Sabuco J, Sanju\'an MAF.
 How to minimize the control frequency to sustain transient chaos using partial control.
 {\em Commun. Nonlinear Sci. Numer. Simul.} 2014; 19:726-737.


\end{thebibliography}
\end{document}